\documentclass{blois}

\usepackage{units}
\def\Journal#1#2#3#4{{#1} {\bf #2}, #3 (#4)}

\usepackage{xspace}
\usepackage{siunitx}
\newcommand{\pt}{\ensuremath{p_{\mathrm{T}}}\xspace}
\newcommand{\GeV}{\unit{GeV}}
\newcommand{\TeV}{\unit{TeV}}
\newcommand{\PW}{\ensuremath{\mathrm{W}}\xspace}
\newcommand{\PZ}{\ensuremath{\mathrm{Z}}\xspace}
\newcommand{\WZ}{\ensuremath{\PW\PZ}\xspace}
\newcommand{\cPZ}{\ensuremath{\mathrm{Z}}\xspace}
\newcommand{\Pp}{\ensuremath{\mathrm{p}}\xspace}

\newcommand{\ptmiss}{\ensuremath{p_{\mathrm{T}}^{\mathrm{miss}}}\xspace}
\newcommand{\MET}{\ptmiss}
\newcommand{\cPq}{\ensuremath{\mathrm{q}}\xspace}
\newcommand{\cPaq}{\ensuremath{\mathrm{\bar{q}}}\xspace}
\newcommand{\tth}{\ensuremath{\mathrm{t\bar{t}H}}\xspace}
\newcommand{\ttv}{\ensuremath{\mathrm{t\bar{t}V}}\xspace}
\newcommand{\ttbar}{\ensuremath{\mathrm{t\bar{t}}}\xspace}
\newcommand{\cPqb}{\ensuremath{\mathrm{b}}\xspace}

\newcommand{\thetaW}{\ensuremath{\theta^{\PW}}}
\newcommand{\thetaZ}{\ensuremath{\theta^{\PZ}}}

\def\PRL{\em Phys. Rev. Lett.}

\def\be{\begin{equation}}
\def\ee{\end{equation}}
\def\bea{\begin{eqnarray}}
\def\eea{\end{eqnarray}}

\newcommand{\Photo}{\includegraphics[height=35mm]{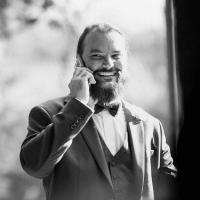}}

\begin{document}
\vspace*{4cm}
\title{Electroweak Precision Measurements in Diboson Production at CMS}

\author{ Pietro Vischia\\ (on behalf of the CMS Collaboration) }

\address{ Centre for Cosmology, Particle Physics and Phenomenology - CP3, Université catholique de Louvain, 2 Chemin du Cyclotron - Box L7.01.05, B-1348 Louvain-la-Neuve, Belgium}

\maketitle\abstracts{
In this contribution, I have outlined recent precision measurements of the standard model (SM) multiboson production at CMS.
A study of diboson production at 5~\TeV constitutes an important probe of the SM at a new energy, and the data favour NNLO predictions obtained by \textsc{MATRIX}.
A study of \WZ production at 13~\TeV constitutes the most comprehensive study of \WZ production to date, containing inclusive and differential cross section measurements,
charge asymmetry measurements, constraints on the LHC proton parton distribution functions, and constraints on anomalous values of the $\PW\PW\PZ$ trilinear gauge coupling.
No evidence for new physics is found, and all the results favour SM predictions calculated at NNLO using \textsc{MATRIX}.
}

\section{Multiboson Production at CMS}
The associated production of two or three vector bosons (\PW or \PZ) at the LHC constitutes a set of important processes that help shed light on the standard model (SM) of particle physics.
On one side, multiboson production constitutes an important background to measurements of the Higgs boson properties or to searches of new physics.
On the other side new physics may appear directly in multiboson production in the form of anomalous values of the triple and quadruple gauge bosons couplings that intervene in the production of these processes.
In this Manuscript, I will describe the most recent precision measurement of SM multiboson production performed by the CMS Collaboration with LHC data, namely the study of diboson production at a centre-of-mass energy of 5~\TeV~\cite{vv5tev} and the study of electroweak (EWK) \WZ production at 13~\TeV~\cite{wz}. Associated WZ production via vector boson scattering is the focus of the contributions by Oleg Kuprash and Mattia Lizzo at this very same conference.

\section{Diboson production at 5~\TeV}
\begin{figure}
\begin{minipage}{0.49\linewidth}
\centerline{\includegraphics[width=0.9\linewidth]{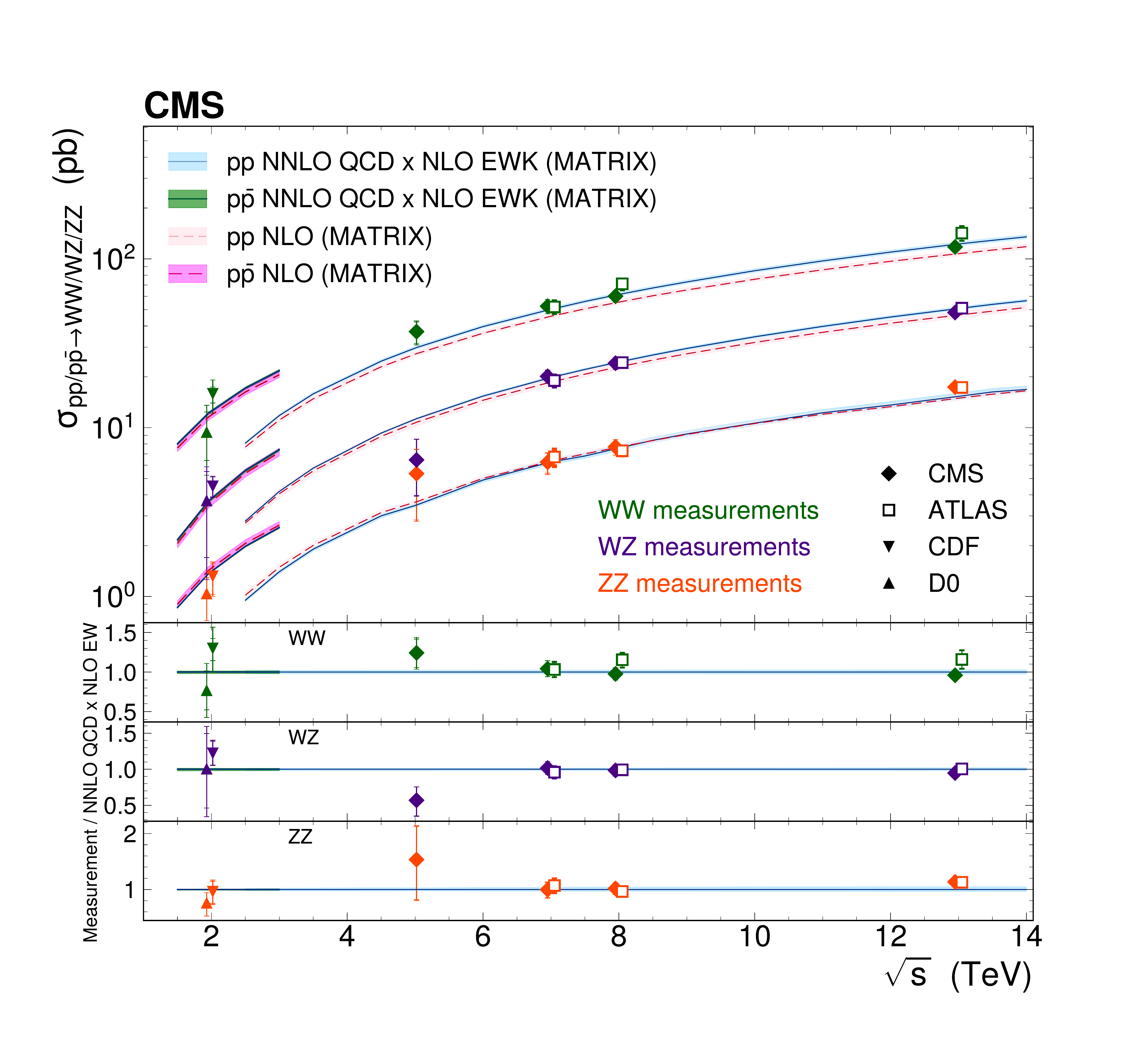}}
\end{minipage}
\hfill
\begin{minipage}{0.49\linewidth}
  \centerline{\includegraphics[width=0.9\linewidth]{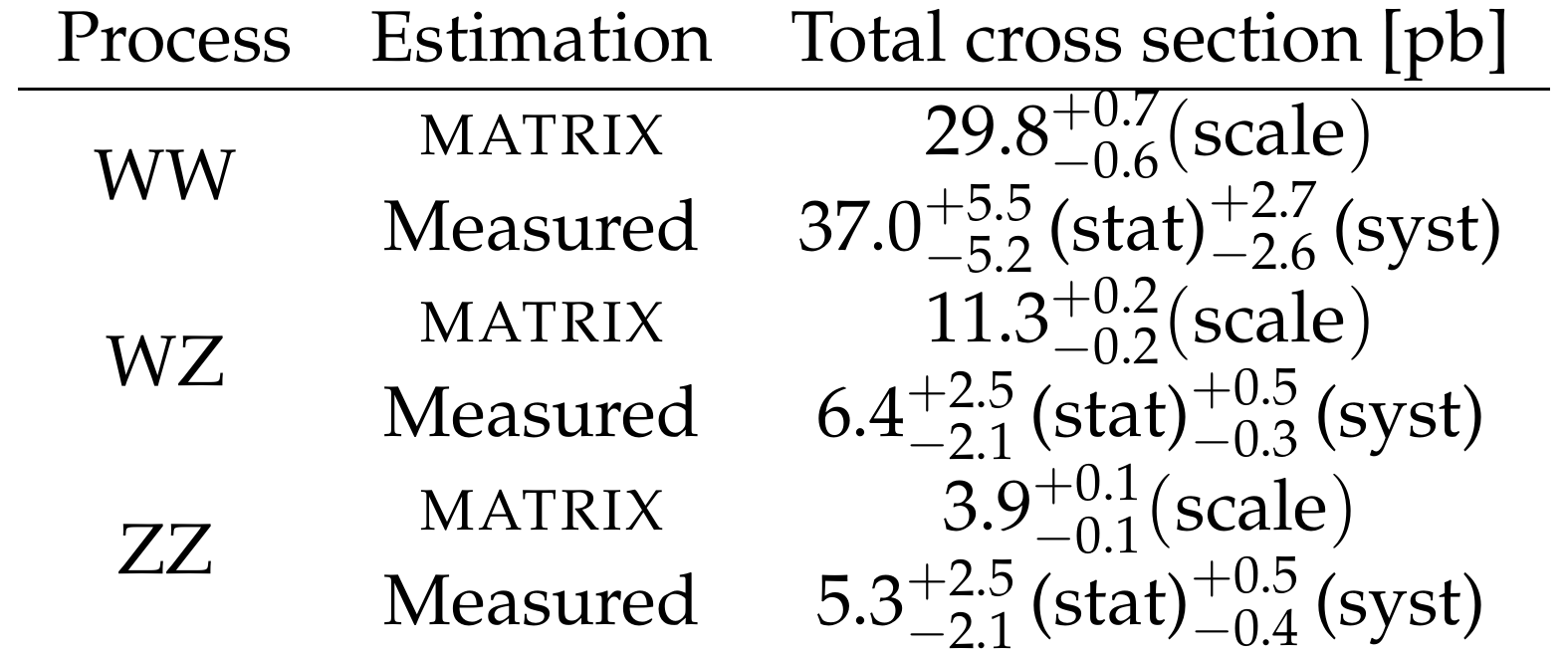}}
\end{minipage}
\caption[]{Evolution of VV cross section measurements as a function of the centre-of-mass energy (left). Measured cross sections for VV production, compared with predictions from \textsc{MATRIX}~\cite{mat1} (right). Figure and table reproduced from the CMS paper~\cite{vv5tev}.}
\label{fig:5TeV}
\end{figure}

Diboson ($\PW\PW$, \WZ, and $\PZ\PZ$) production constitutes an important probe for the dependence of SM cross sections on the beam energy; measuring diboson production cross sections at a center-of-mass energy of 5~\TeV{} is therefore of paramount importance.
The CMS analysis~\cite{vv5tev} makes use of single-lepton triggers to study diboson production in a final state characterized by the presence of multiple leptons: $\PW\PW$ production is studied by requiring two opposite-charge different-flavour leptons, plus transverse mass requirements and vetoing jet production; $\PW\cPZ$ production is targeted in a three-lepton and in a two-muons same-sign region with additional requirements that identify the pairs of leptons that most likely originate from the decay of a \PZ{} boson; and the $\PZ\PZ$ is studied in a four-lepton region and in a two-lepton opposite-charge same-flavour requiring the missing transverse energy \ptmiss to be $\ptmiss>50\GeV$. All the background contributions are estimated from simulation, except for the contribution from objects misidentified as leptons (\textit{nonprompt lepton} contribution), which is estimated from data. The results, illustrated and tabulated in Fig.~\ref{fig:5TeV}, are well in agreement with NNLO predictions from \textsc{MATRIX}~\cite{mat1}, except in \WZ, where the agreement is within two standard deviations.

\section{WZ production at 13 TeV}
\begin{figure}
\begin{minipage}{0.32\linewidth}
\centerline{\includegraphics[width=0.9\linewidth]{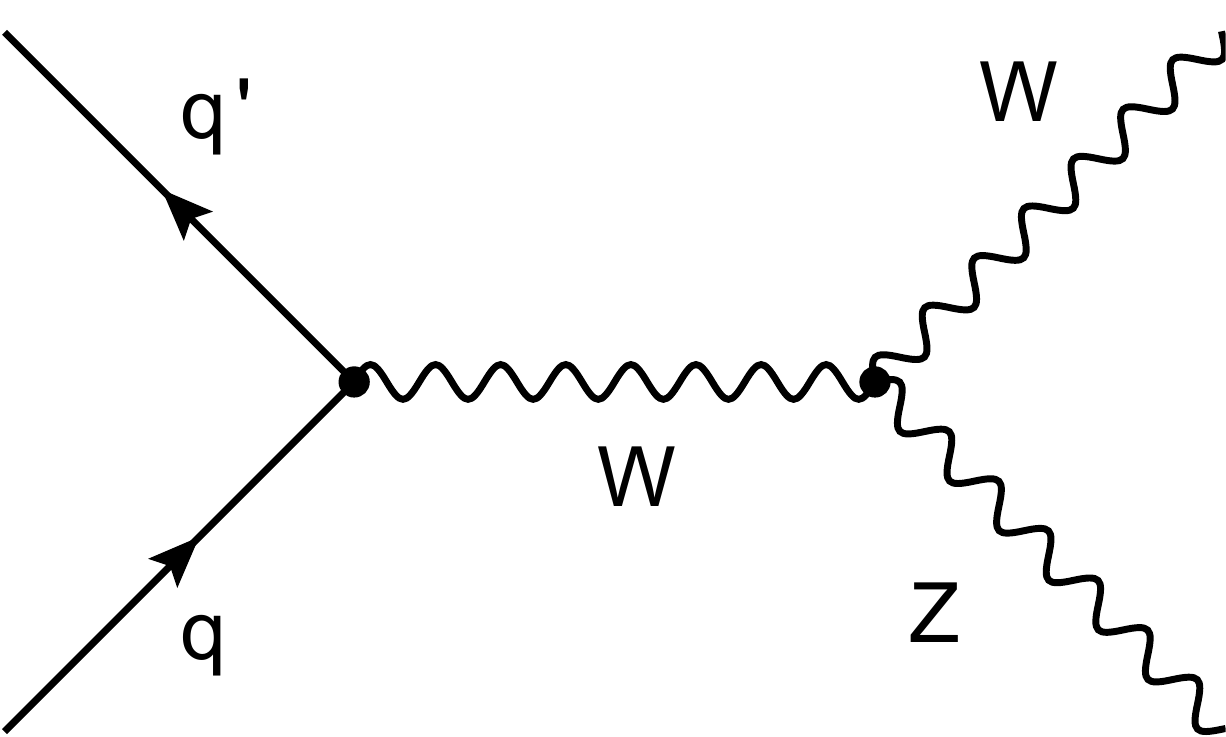}}
\end{minipage}
\hfill
\begin{minipage}{0.32\linewidth}
  \centerline{\includegraphics[width=0.9\linewidth]{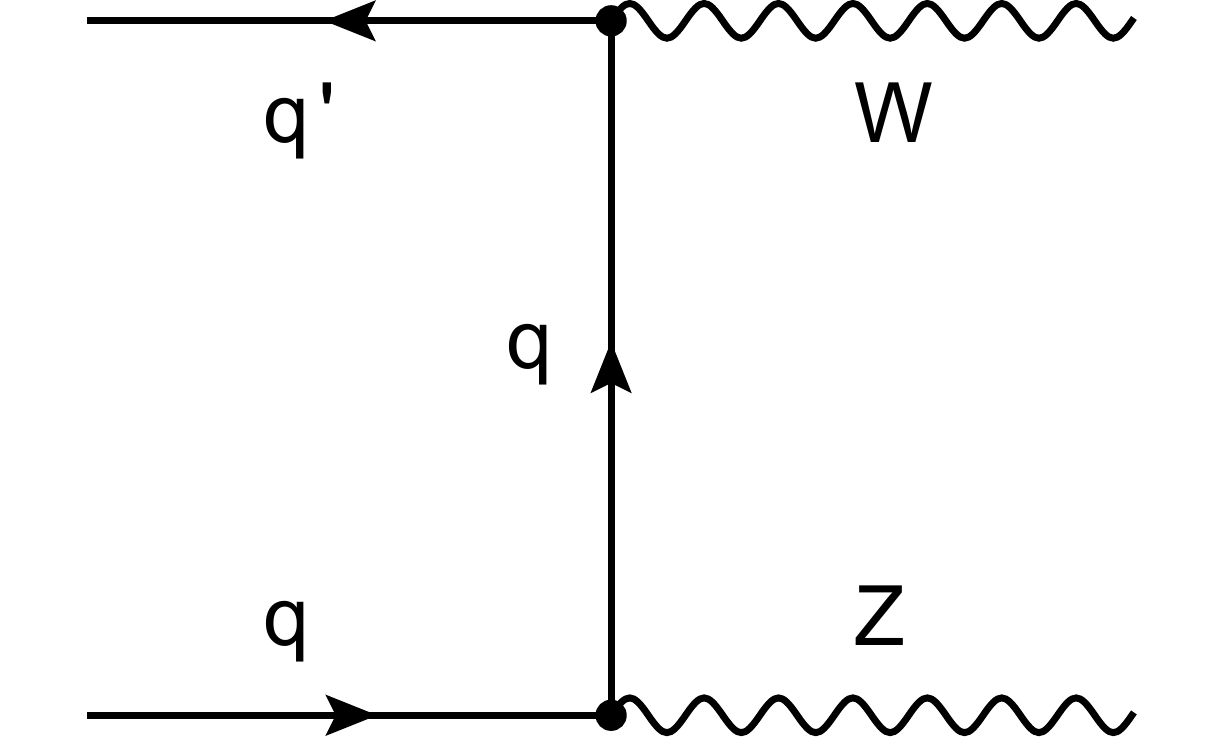}}
\end{minipage}
\hfill
\begin{minipage}{0.32\linewidth}
  \centerline{\includegraphics[width=0.9\linewidth]{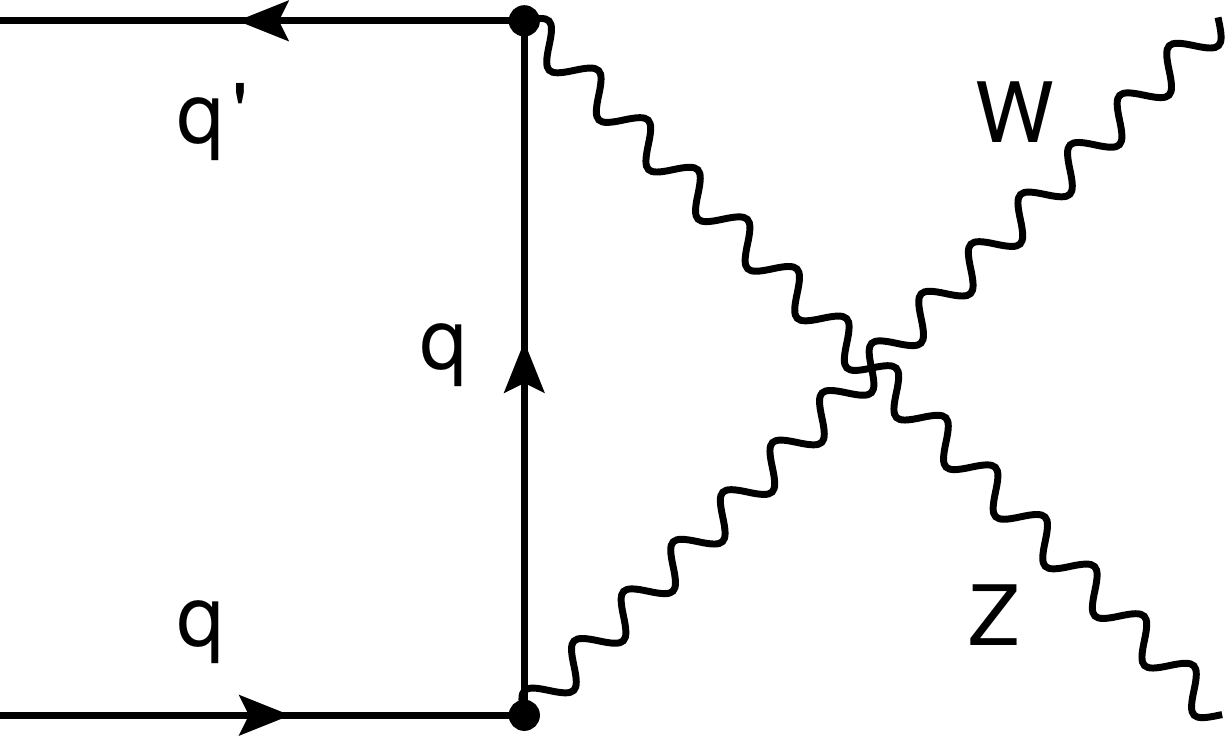}}
\end{minipage}
\caption[]{Feynman diagrams for \WZ production. The diagram on the left is sensitive to anomalous values of the $\PW\PW\PZ$ coupling. Figures reproduced from the CMS paper~\cite{wz}.}
\label{fig:wzdiagrams}
\end{figure}

The associated production of a \PW and a \PZ boson at 13 TeV has been studied using the full CMS Run II data set~\cite{wz}.
This process features a charged final state that is sensitive to the quark parton distribution functions (PDFs)~\cite{pdfs}, because at the tree level it is completely dominated by $\cPq\cPaq'$ states.
\WZ is also sensitive at tree level to anomalous values of the WWZ triple gauge coupling, and it is the dominant SM background process in any analysis targeting
trilepton final states with low hadronic activity, such as \tth~\cite{tth_multilepton_legacy}, \ttv (the first full Run~2 determination of $\ttbar\PW$ cross section has been published in ~\cite{tth_multilepton_legacy}), supersymmetric electroweak searches~\cite{ewkino}, etc. The diagrams for \WZ production are shown in Fig.~\ref{fig:wzdiagrams}.

This CMS study~\cite{wz} is a comprehensive study of \WZ production, leveraging the multilepton final state to measure fiducial, inclusive, and differential cross sections, vector boson polarization, and the value of the trilinear gauge coupling that intervenes in Fig.~\ref{fig:wzdiagrams} (left). For the first time, \WZ production is used to probe CP-violating parameters in effective field theory (EFT) rather than the traditional CP-conserving ones, and we use Bayesian methods to innovatively constrain the proton PDFs predicted by the LHAPDF Collaboration~\cite{pdfs}.

The analysis uses a set of single-lepton and dilepton triggers with an overall efficiency of about 100\%, and with respect to previous publications~\cite{wz2016}, we have improved the lepton reconstruction to provide additional background suppression. All leptons are \textit{dressed}, i.e. the photon momenta in a radius of 0.10 in the $(\eta, \phi)$ plane are added to the lepton momentum, and the dedicated boosted decision tree that identifies leptons has been now retuned based on optimizations we performed in the context of the CMS Observation of ttH production~\cite{tth_observation}. For electrons, a tight-charge criterium is also included: it reduces the acceptance by $<1\%$ while reducing the background contribution from \textit{charge flips}, i.e. misreconstructed electron charge, from 0.2\% to 0.03\%.
We compute lepton identification efficiencies~\cite{carlos} in simulated events, using \WZ production as a signal, and the set of all nonprompt background contributions---mostly \ttbar and Drell-Yan---as background, according to the formula:
\begin{equation}
  \epsilon(MVA) := \frac{\#\text{events in SR(Loose + MVA cut)}}{\#\text{events in SR(loose)}}\,.
\end{equation}

The event selection consists of a baseline selection of three light leptons (electrons or muons), out of which an opposite-sign same-flavour pair compatible with the hypothesis of coming from a \PZ boson decay is required. A no-heavy-flavour-activity, high-trilepton-mass signal region (SR) is used in conjunction with three control regions (CRs) where we invert some of the signal region requirements to target specific backgrounds: a $\PZ\PZ$ CR is characterized by requiring four leptons, a Conversions CR by requiring low \ptmiss and dilepton mass, and a \ttbar CR by requiring the presence of jets coming from the fragmentation of \cPqb quarks and no \PZ mass window requirement. We compute the fiducial cross sections by means of a maximum likelihood fit to the lepton flavour composition in each of the regions while also constraining the normalization of the three backgrounds in their CRs. For the $\PZ\PZ$ region, the lepton flavour is taken as the flavour of the three leading-\pt leptons, to mimic what would affect us in the SR if we didn't reconstruct the fourth lepton. 

\begin{figure}
\begin{minipage}{0.32\linewidth}
\centerline{\includegraphics[width=0.9\linewidth]{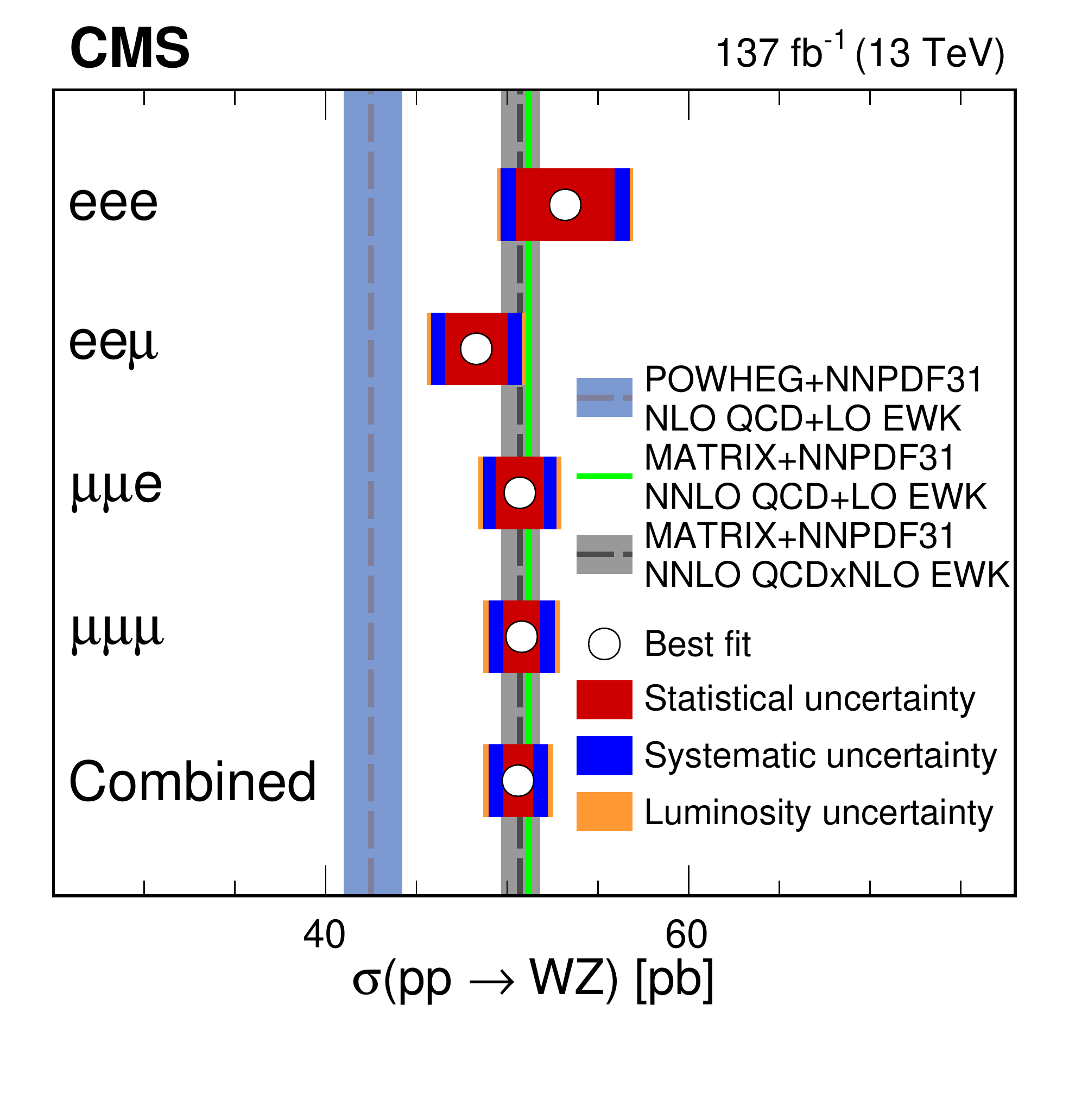}}
\end{minipage}
\hfill
\begin{minipage}{0.32\linewidth}
  \centerline{\includegraphics[width=0.9\linewidth]{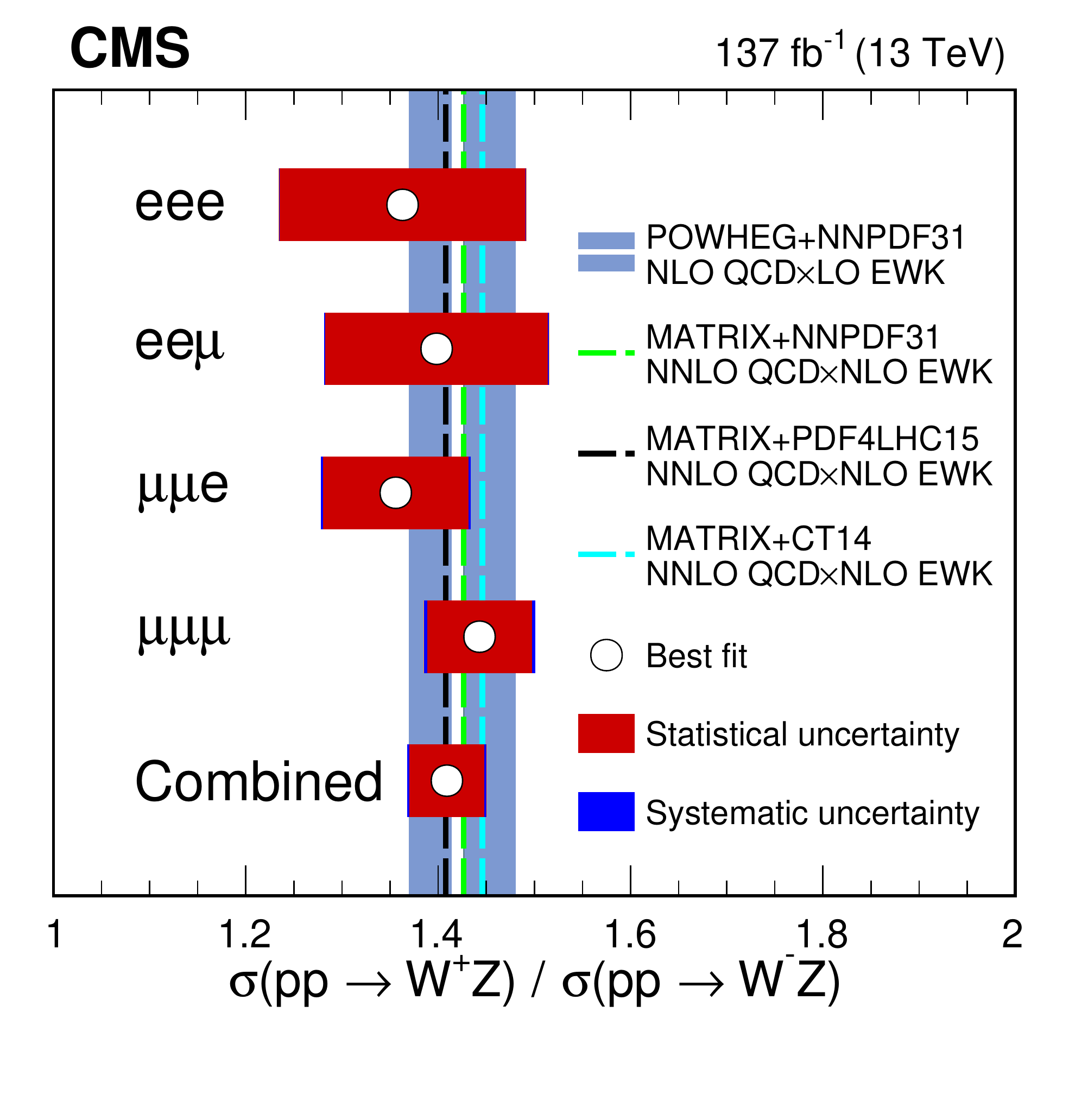}}
\end{minipage}
\hfill
\begin{minipage}{0.32\linewidth}
  \centerline{\includegraphics[width=0.9\linewidth]{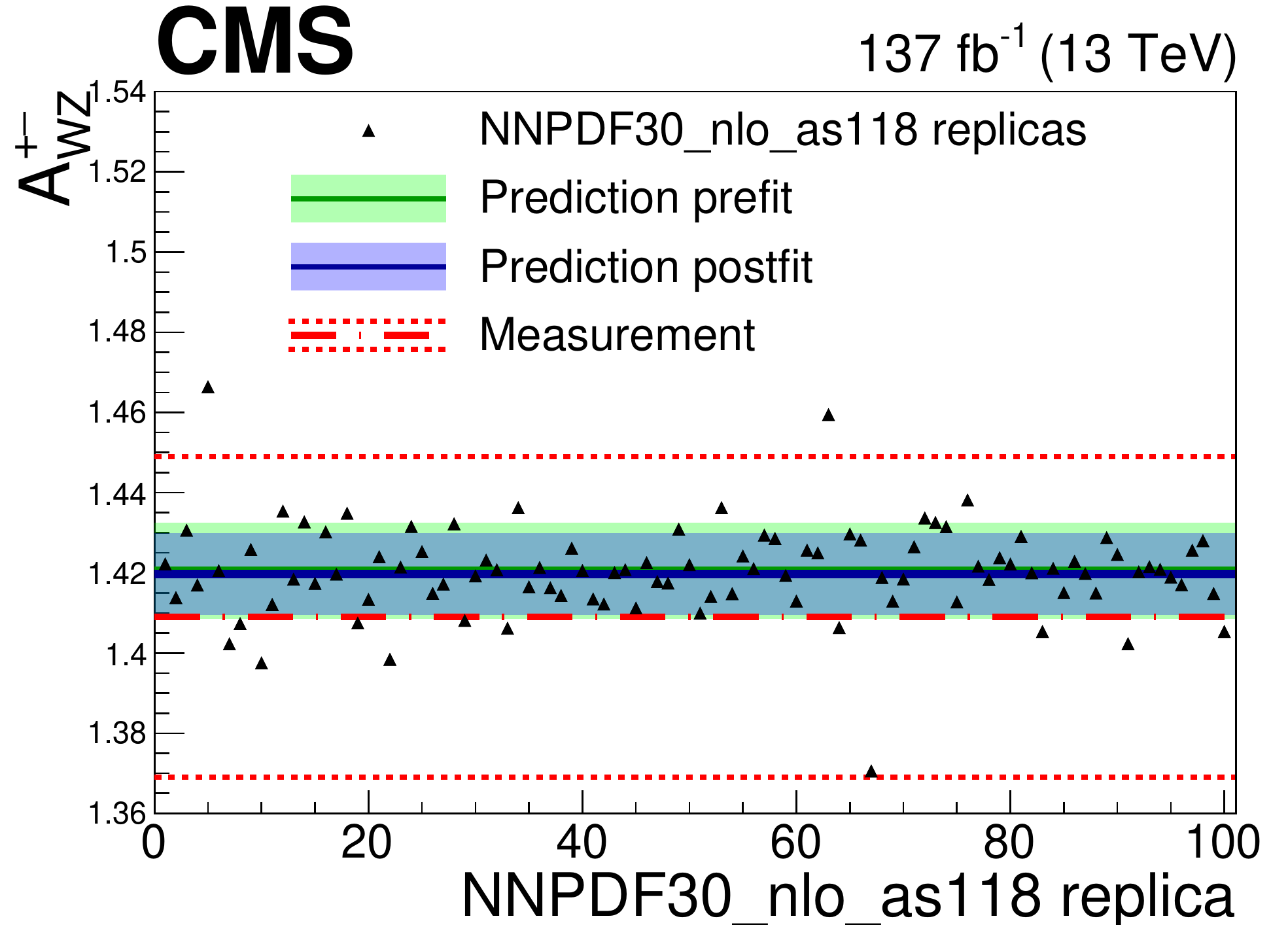}}
\end{minipage}
\caption[]{Cross section measurements for \WZ production (left) and charge asymmetry (middle). Constraints on the proton PDFs~\cite{pdfs} uncertainty obtained via Bayesian reweighting (right). Figures reproduced from the CMS paper~\cite{wz}.}
\label{fig:xsecs}
\end{figure}

We find that the observed data favour the NNLO predictions from \textsc{MATRIX}~\cite{mat1}, and we extrapolate the fiducial results to the inclusive phase space accounting for leptonic
branching fractions. Our results, displayed in Fig.~\ref{fig:xsecs} and tabulated in the paper~\cite{wz}, are affected by a 4\% overall uncertainty on each lepton flavour final state, which is smaller than the best determination to date (5\% by the ATLAS Collaboration~\cite{atlas}). The dominant systematic uncertainty components are, in decreasing order, the beam luminosity, the \cPqb{} quark tagging, and the nonprompt lepton identification, whereas the statistical uncertainty accounts for about 90\% of the total uncertainty.
Combining the result across the four lepton flavour final states yields a cross section measured with an overall uncertainty of 3\%, which is about half of similar measurements by ATLAS~\cite{atlas} and CMS~\cite{wz2016} and is below the theoretical uncertainties in the \textsc{POWHEG}~\cite{pow1,pow2,pow3,pow4,pow5} predictions. The results are illustrated in Fig.~\ref{fig:xsecs} (left, middle), also for the ratio between charge asymmetry ratio
\begin{equation}
  A^{+-}(\mathrm{WZ}) = \frac{\sigma_{\mathrm{fid}}(\Pp\Pp \rightarrow \PW^+\cPZ)}{\sigma_{\mathrm{fid}}(\Pp\Pp \rightarrow \PW^-\cPZ)}\,.
\end{equation}
The asymmetry ratio is a consequence of the asymmetry in the up and down quark proton PDFs~\cite{pdfs}. We therefore use our measurement to compute a Bayesian posterior predictions for the LHC proton PDFs~\cite{pdfs}, constraining the uncertainty in their prediction by about 10\% with respect to the theoretical value, as illustrated in Fig.~\ref{fig:xsecs} (right).

The helicity of the \PW boson has been measured in several production modes, such as \ttbar, single top, and \PW+jets, and by ATLAS in \WZ production~\cite{atlas}.
A measurement in \WZ production was difficult so far because of the limited amount of data events available in the \WZ signal regions.
With the full Run~2 dataset, we have been able to measure the polarization of both the \PW and the \PZ boson~\cite{pol},
by relying on assigning each lepton to its parent boson (which we are able to do with about 95\% efficiency),
and reconstructing the angle between the parent boson and the resulting lepton, based on templates from simulation. At leading order and assuming no interference, we express the polarization angles as:
\begin{equation}
  \frac{1}{\sigma}\frac{\sigma}{\cos{\thetaW{}^{\pm}}} = \frac{3}{8} \left\{ \bigl[1 \mp \cos(\thetaW{}^{\pm})\bigr]^2 f_{\mathrm{L}}^{\PW} + \bigl[1 \pm \cos(\thetaW{}^{\pm})\bigr]^2 f_{\mathrm{R}}^{\PW} + 2 \sin^2(\thetaW{}^{\pm}) f_{0}^{\PW} \right\}
\end{equation}
and
\begin{equation}
  \frac{1}{\sigma}\frac{\sigma}{\cos{\thetaZ}} = \frac{3}{8} \left\{ \bigl[1 +  \cos^2(\thetaZ) - 2c\cos(\thetaZ)\bigr] f_{L}^{\PZ} + \bigl[1 +  \cos^2(\thetaZ) + 2c\cos(\thetaZ)\bigr]  f_{R}^{\PZ} + 2 \sin^2(\thetaZ) f_{0}^{\PZ}\right\}\,,
\end{equation}
as a function of the three polarization fractions in the helicity frame. A corresponding equation governs the polarization angle for the \PZ boson.
A maximum likelihood fit of the polarization angle is performed to determine the three fractions and the overall \WZ normalization, with the constraint that the fractions must sum up to unity.
Our paper~\cite{wz} reports the results in tabular and graphical forms for one-, two-, and three-dimensional fits were the remaining free parameters are set to their SM prediction. The results are generally in agreement with predictions from \textsc{POWHEG}~\cite{pow1,pow2,pow3,pow4,pow5} and \textsc{MATRIX}~\cite{mat1}.

Differential cross sections are measured as a function of several observables: besides those already employed in our previous paper~\cite{wz2016}, we also probed the polarization angles of the \PW and \PZ bosons, and the jet multiplicity (which is a probe for the validity of the jet simulation. The measurement generally favour the predictions from \textsc{MATRIX}~\cite{mat1,mat2}, and are illustrated and tabulated~\cite{wz} both in merged form and split by leptonic charge and flavour.

\begin{figure}
\begin{minipage}{0.32\linewidth}
\centerline{\includegraphics[width=0.9\linewidth]{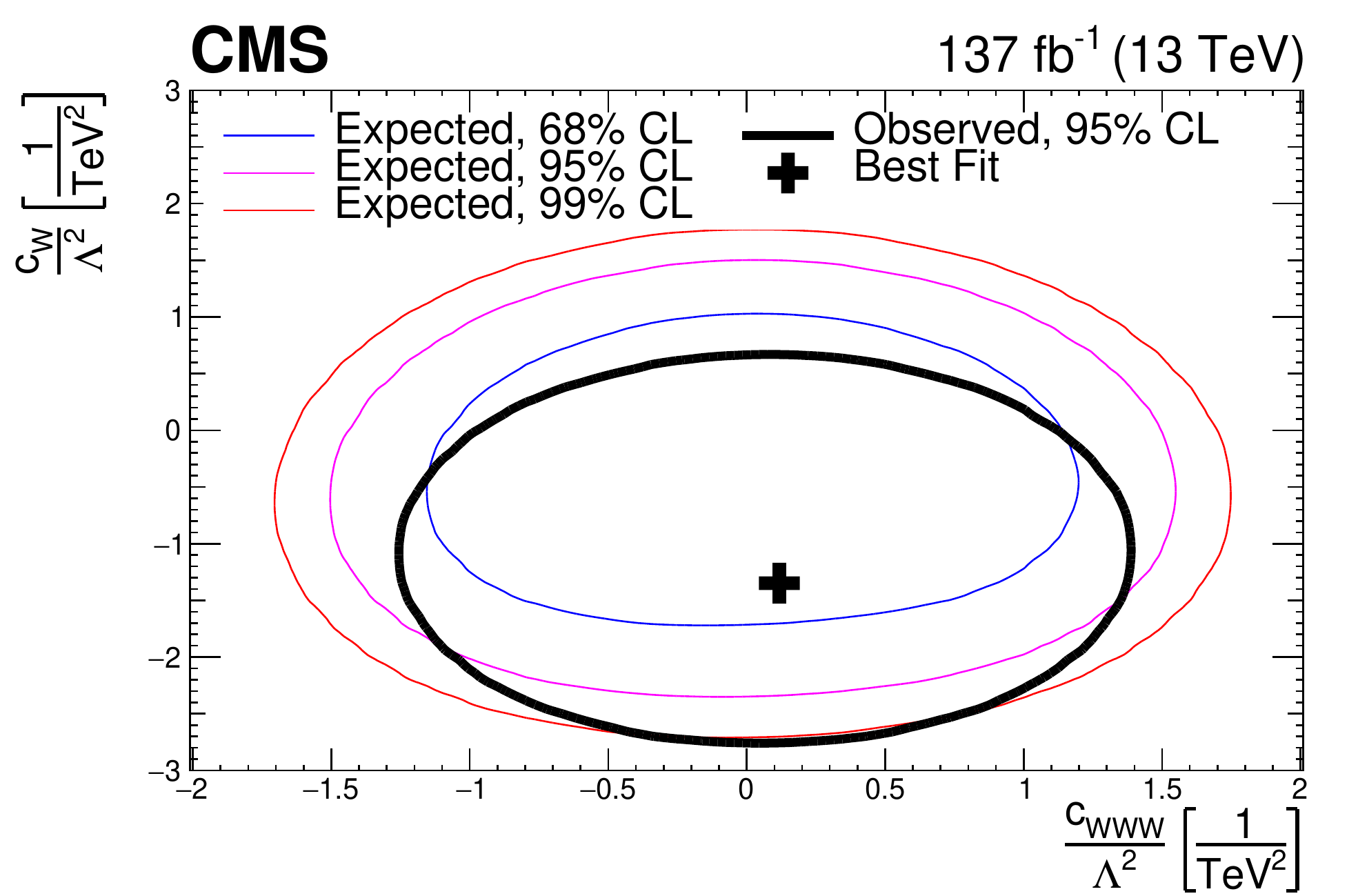}}
\end{minipage}
\hfill
\begin{minipage}{0.32\linewidth}
  \centerline{\includegraphics[width=0.9\linewidth]{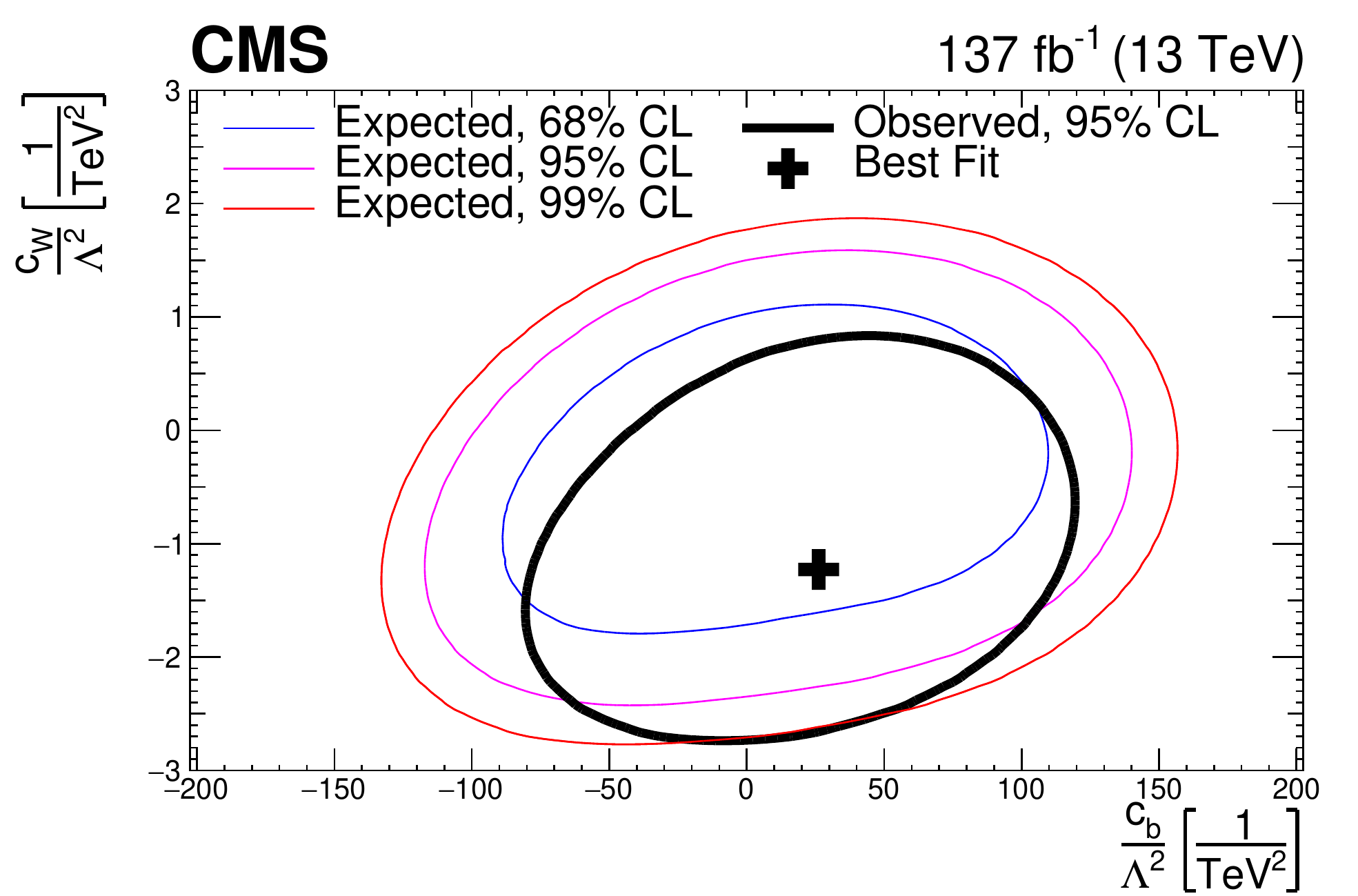}}
\end{minipage}
\hfill
\begin{minipage}{0.32\linewidth}
  \centerline{\includegraphics[width=0.9\linewidth]{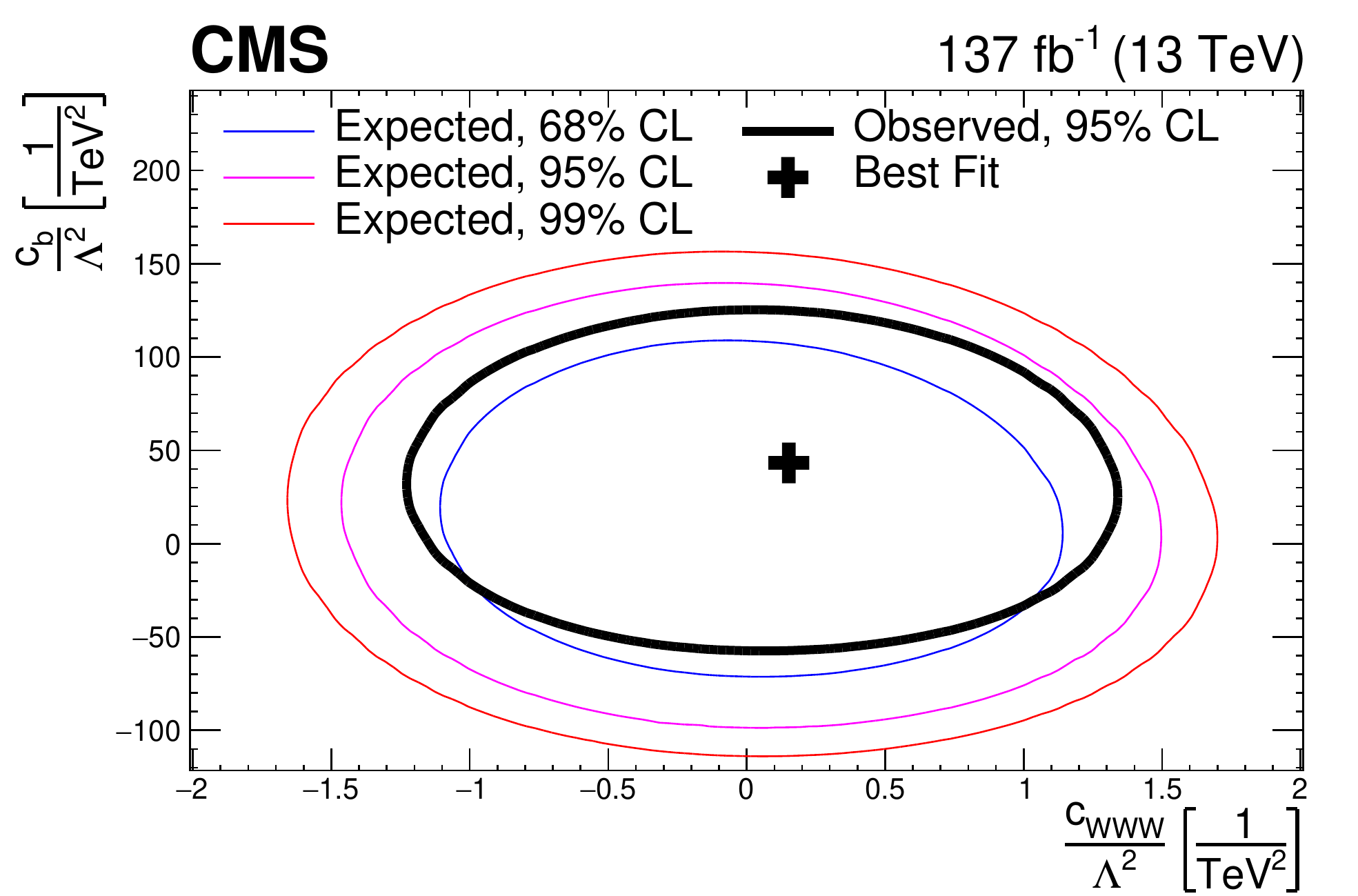}}
\end{minipage}
\caption[]{Two-dimensional confidence regions for pairs of couplings in the helicity frame. Correlations between the parameters can be inferred from the elliptical shape and inclination of the regions. Figures reproduced from the CMS paper~\cite{wz}.}
\label{fig:couplings}
\end{figure}

Finally, constraints on anomalous values of the $\PW\PW\PZ$ trilinear gauge coupling are set. \WZ production is sensitive to multiple BSM effects as effective low energy theories.
We use a generic model with three couplings, that the SM predicts to have the values $g_1^Z=1$, $k_Z=1$, $\lambda_Z=0$. We are less sensitive to the $k_Z$ term, because in \WZ production we have no access to the \pt of the \PW~propagator. Deviations from SM values of the couplings are visible at high \pt, and we therefore set constraints on the couplings by performing a maximum likelihood fit of the $M_{\ell\ell\ell\MET}$ distribution: the sensitivity to EFT effects comes from the tails of that distribution. One-, two-, and three dimensional confidence regions (fixing to their SM predictions the parameters that are not determined in each fit) are tabulated and illustrated in the paper~\cite{wz}. We find to evidence for anomalous values of the couplings. Small correlations between the EFT parameters are inferred from the two-dimensional plots shown in Fig.~\ref{fig:couplings}. The paper~\cite{wz} also contains the results converted to the Warsaw basis.

\begin{figure}
\centerline{\includegraphics[width=0.49\linewidth]{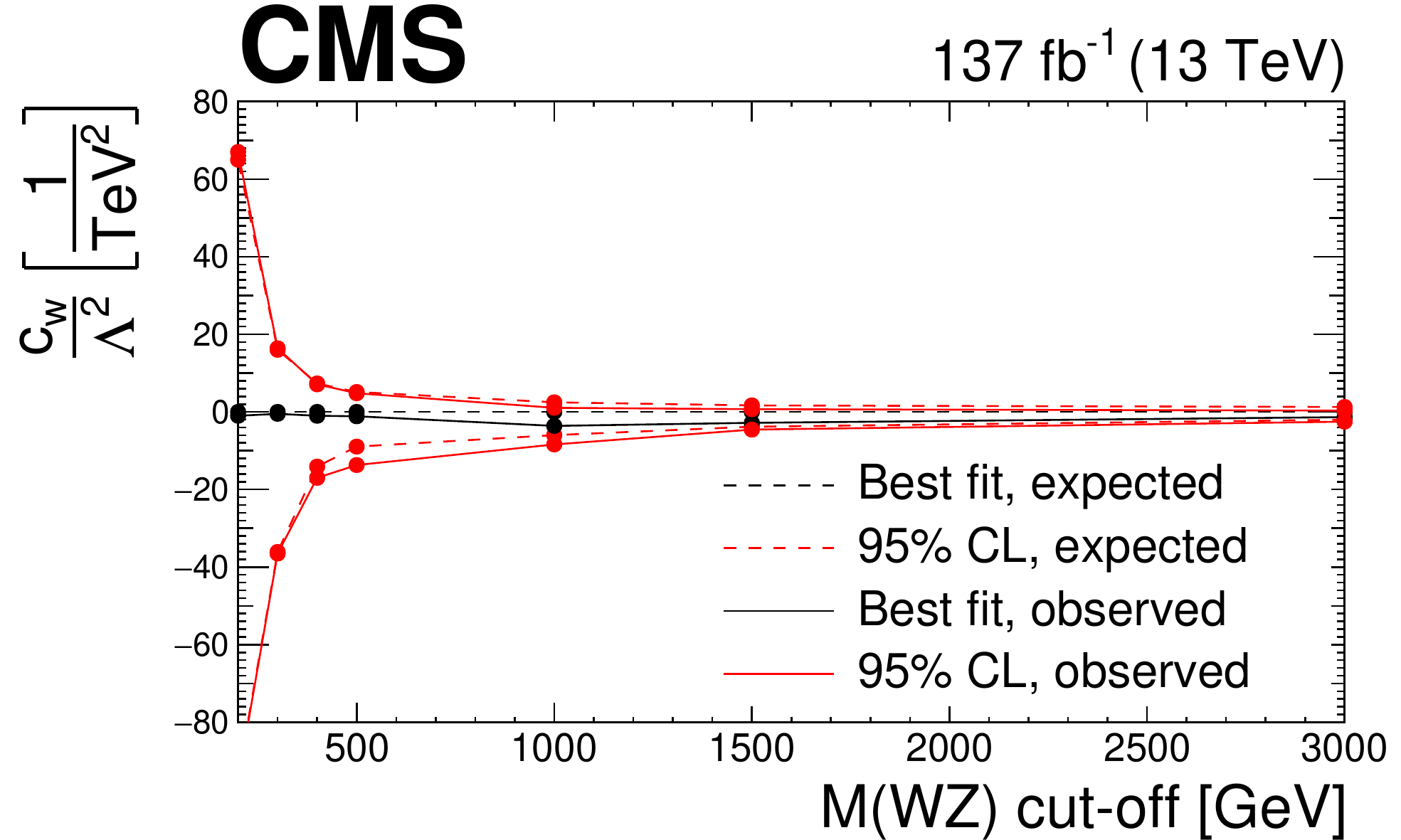}}
\caption[]{Example evolution of the confidence regions on one EFT parameter as a function of the cutoff scale for EFT suppression at high energies. Figures reproduced from the CMS paper~\cite{wz}.}
\label{fig:cutoff}
\end{figure}

In EFTs for \WZ production, interference terms between the SM and beyond-SM physics are described in dimension six operators by linear terms that are $\Lambda^{-2}$- and $\Lambda^{-4}$-suppressed.
Dimension-eight operators would introduce an additional interference term, and the results would be accurate only up to $\Lambda^{-2}$-suppressed contributions.
We performed additional maximum likelihood fits by using only the linear terms of the quadratic fitting functions, to check the effect of dropping $\Lambda^{-4}$ terms from the modelling.
This is an important cross check of the structure of EFT theories, and the results are illustrated and tabulated in the paper~\cite{wz}.
Finally, dimension-6 operators lead to nonphysical results characterized by unitarity breaking leading to cross section values of infinity at arbitrarily high energies.
We introduce a cutoff scale to suppress EFT at high energies, and show the evolution of the confidence regions for the EFT parameters as a function of the cutoff in Fig.~\ref{fig:cutoff}.

\section{Summary}
In this contribution, I have outlined recent precision measurements of the standard model (SM) multiboson production at CMS.
A study of diboson production at 5~\TeV~\cite{vv5tev} constitutes an important probe of the SM at a new energy, and the data favour NNLO predictions obtained by \textsc{MATRIX}~\cite{mat1}.
A study of \WZ production at 13~\TeV~\cite{wz} constitutes the most comprehensive study of \WZ production to date, containing inclusive and differential cross section measurements,
charge asymmetry measurements, constraints on the LHC proton PDFs~\cite{pdfs}, and constraints on anomalous values of the $\PW\PW\PZ$ trilinear gauge coupling.
No evidence for new physics is found, and all the results favour SM predictions calculated at NNLO using \textsc{MATRIX}~\cite{mat1,mat2}.

\end{document}